**Interactions between β-endorphin and kisspeptin neurons of the ewe arcuate nucleus are modulated by photoperiod.**

Vincent Hellier[1], Hugues Dardente[1], Didier Lomet[1], Juliette Cognié[1], Laurence Dufourny[1]

[1] CNRS, IFCE, INRAE, Université de Tours, PRC, F-37380, Nouzilly, France

*All correspondence should be addressed to:

>Dr Laurence Dufourny
>
>UMR PRC
>
>Centre INRAE Val de Loire
>
>37380 Nouzilly
>
>FRANCE
>
>laurence.dufourny@inrae.fr
>
>Phone : (+33) 247427903

Short title: β-endorphin and kisspeptin interactions in the ewe

**Conflict of interest:** none

Keywords : kisspeptin, opioids, mediobasal hypothalamus, seasonal reproduction, ewe.

**Abstract**

Opioid peptides are well-known modulators of the central control of reproduction. Among them, dynorphin coexpressed in kisspeptin (KP) neurons of the arcuate nucleus (ARC) has been thoroughly studied for its autocrine effect on KP release through κ opioid receptors. Other studies have suggested a role for β-endorphin (BEND), a peptide cleaved from the proopiomelanocortin (POMC) precursor, on food intake and central control of reproduction. Similarly to KP, BEND content in the ARC of sheep is modulated by day length and BEND modulates food intake in a dose-dependent manner. As KP levels in the ARC vary with photoperiodic and metabolic status, a photoperiod-driven influence of BEND neurons on neighboring KP neurons is plausible.

The aim of this study was therefore to investigate a possible modulatory action of BEND on KP neurons located in the ovine ARC. Using confocal microscopy, numerous KP appositions on BEND neurons were found but there was no photoperiodic variation in the number of these interactions in ovariectomized, estradiol-replaced ewes. In contrast, BEND terminals on KP neurons were twice as numerous under short days (SD), in ewes having an activated gonadotropic axis, as compared to anestrus ewes under long days (LD). Injection of 5µg BEND into the third ventricle of SD ewes induced a significant and specific increase of activated KP neurons (16% versus 9% in controls) while the percentage of overall activated (c-Fos positive) neurons, was similar between both groups. These data suggest a photoperiod-dependent influence of BEND on KP neurons of the ARC, which may influence GnRH pulsatile secretion and inform KP neurons on the metabolic status.

**Introduction**

Mammalian species have developed a wide range of strategies to adapt to their changing environment hence reinforcing species survival. Among them, seasonal reproduction resulting from the neurochemical translation of day length into melatonin secretion at night allows for the birth of pups at the most suitable moment of the year in terms of food availability and climate. In domestic seasonal species such as horse, goat and sheep, births occur in spring under natural conditions and several strategies have been developed to master reproduction around the year. These latter include the association of light treatments, metabolic flushing and use of exogenous hormones. Data gathered in ewes over two consecutive years have shown that food-restricted ewes had an anestrus season that started earlier and lasted longer than well fed animals (1). Similar results obtained in mares also pinpointed toward a modulatory influence of the metabolic status on the duration of anestrus season (2). It appears therefore that metabolic cues are important modulators of the duration of the breeding season. Several orexigenic and anorexigenic neuronal populations have been described in the hypothalamus and more specifically in the arcuate nucleus (ARC) including neuropeptide Y neurons and proopiomelanocortin (POMC) neurons (for review see:(3)). POMC is cleaved into several different peptides among which β-endorphin (BEND), a 31-amino acid peptide that exerts both orexigenic and anorexigenic effects, depending on the dose injected (4, 5). Furthermore, BEND secretion is increased in long photoperiod (LD), which is associated with seasonal anestrus in sheep (6). We and others used sheep to reveal significant photoperiod changes in the number of BEND terminals on GnRH neurons (7) and of PSA-NCAM expression on BEND neuronal somas in the ARC caudal two-thirds (8), which stress the role of synaptic plasticity in the central control of breeding. Moreover, intracerebroventricular (icv) infusion of 2 or 10 µg BEND in follicular phase ewes resulted in a prolonged inhibition of LH pulse secretion while intravenous injection of 50µg BEND failed to modify LH levels, which suggests that BEND acts at the hypothalamic level rather than at neuronal terminals in the median eminence (9). BEND binds µ opioid receptors with high affinity (pKi=9 in rats) and δ and κ opioid receptors with lower affinities (pKi=8.3 and pKi=6.3-7.9 in humans respectively; IUPHAR website). However δ opioid receptors appear to be scarcely distributed in the human and rodent hypothalamus (for review see: (10)) and only κ and µ opioid receptors appear to be involved in the central control of reproduction in ewes (11). BEND neurons express sex steroid receptors (12, 13) and project toward the organum vasculosum of the lamina terminalis (OVLT) where most GnRH neurons are located (13). Therefore, BEND may transduce steroid feedback to GnRH neurons expressing µ receptors, at least in rodents (14).

The central control of reproduction depends on the timely release of GnRH from neurons themselves highly dependent upon the secretion of kisspeptin (KP), an amidated peptide produced in neurons located in two populations found in the preoptic area and ARC in ewes (15). The prominent role of KP on GnRH release has been widely established since the early 2000's as individuals with a mutation in KP or its receptor, Kiss1r, display hypogonadotropic hypogonadism resulting in infertility (16, 17). Most GnRH neurons express Kiss1r (18) and central infusion of KP-10, the shortest active form of KP, triggers an LH surge in both cyclic and acyclic, out-of-season, ewes (19). KP neurons express sex steroid receptors (15, 20) and relay both positive and negative sex steroid feedbacks to GnRH neurons. Of note, KP synthesis is modulated by photoperiod as short-days (SD) promote an increase of KP levels in the ovine ARC (20-22) and KP terminals on GnRH neurons are more numerous during the breeding season (23). KP neurons are also sensitive to the metabolic status of animals as KP levels in lean ovariectomized ewes are decreased compared to well-fed ewes (24, 25). KP neurons express leptin receptors Ob-R and have reciprocal connections with neuropeptide Y and POMC-expressing neurons within the ARC (24), which suggests a role as metabolic sensors for KP neurons of the ARC. In addition to this metabolic modulation, the activity of KP neurons from the ARC is regulated by neurokinin B, and dynorphin, two peptides co-synthetized within KP neurons, hence their KNDy nickname (26). KNDy neurons express NK3R and κ opioid receptors selectively binding neurokinin B and dynorphin respectively (27, 28). Moreover, KNDy neurons establish reciprocal connections (KNDy-KNDy) that account for GnRH/LH pulsatile release (for review see: (29)) and are considered as the "pulse generator" in several mammalian species including mouse (30) and goat (31). The question remains whether other peptidergic neuronal populations involved in the perception and/or regulation of metabolic status may modulate KP neuron activity. In this respect, BEND neurons appear to be good candidates, at least in sheep.

We hypothesized that BEND inhibitory effect on LH secretion would be mediated through KNDy neurons located in the ARC during short days. Therefore, the objectives of the present study were to assess whether BEND terminals are found in association with KP neurons at the ARC level in ovariectomized estradiol replaced ewes kept either in LD or SD, and to investigate whether the extent of such neuroanatomical relationships vary with gonadotrope axis status. The second aim of this work was to determine if an icv administration of BEND modulates the activation of KP neurons in the ARC of sexually active ewes.

**Material and methods**

*Animals and treatments*

All experimental procedures were approved by the local Ethics Committee (authorization CEEAVDL 2012-09-4) in accordance with local regulations and the approval of the French Ministry of Agriculture. Experiments were conducted on adult 3-year-old Ile-de-France ewes (n=19) obtained from UEPAO (Unité expérimentale PAO n°1297 (EU0028), INRAE Centre Val de Loire, Nouzilly, France). All ewes had free access to food, water, and mineral licks at the INRAE animal facility during breeding and non-breeding seasons. Before the study, all ewes were ovariectomized (OVX) after deep gaseous anesthesia with isoflurane and implanted subcutaneously with a 2-cm Silastic implant containing 17β-estradiol to maintain low physiological serum concentrations of estrogen (approx. 2 pg/mL (32)). In this model, periods of high LH (>1ng/mL) are indicative of an active gonadotropic axis and periods of low LH (<1ng/mL) are associated with gonadotropic axis inactivity (33). Blood samples were performed twice weekly starting from the month preceding the final light treatment transition to the end of the experiment to assess LH levels. Plasma LH concentrations were determined using a double antibody ELISA immunoassay technique published previously (34). The intra- and inter-assay coefficients of variation of the control averaged 8% and 12%, respectively.

*Brain sampling*

All animals were killed by an overdose of pentobarbital (approximately 2.5 g in 5 mL; CEVA Santé Animale, Libourne, France) in a registered slaughterhouse by a certified butcher. For experiments 1 and 2, brains were fixed by perfusion through both ascending carotids with 1 L of 1% sodium nitrite in 0.9% saline, followed by 3 L of fixative containing 4% paraformaldehyde and 15% picric acid in 0.1 M phosphate buffer (PB), pH 7.4. After dissection, brains were soaked in 100 mL of 20% sucrose in 0.1M PB at 4°C. A block containing the whole diencephalon was dissected out, embedded in Tissue-Tek (O.C.T Compound, Zoeterwoude, Netherlands), and quickly frozen using isopentane at -60°C. Brain blocks were stored until sectioning with a cryostat at -20°C (14µm thick frontal sections). Sections were directly collected on 1% 3-aminopropyltriethoxysilane (Sigma, Saint Louis, MO, USA) coated slides. The slides were then stored at -20°C until further processing.

**Experiment 1: Are there neuroanatomical relationships between KP and BEND neurons in the ovine ARC?**

*Light treatment, gonadotrope axis status, and tissue collection*
Before the start of the experiment, all ewes (n=10) were photoperiod-synchronized using an artificial light treatment reproducing long days (LD; 16 hours of light (L): 8 h of darkness (D) per 24h). When LH was <1ng/mL for 3 consecutive samples, 5 ewes were killed. The remaining ewes were then treated with artificial short days (SD; 8L:16D). When LH levels were >1ng/mL for 3 consecutive samples, the 5 remaining ewes were killed. Brain collection was performed as described above.

*KP/BEND immunocytochemistry*

Sections were washed 3 times 5 minutes at room temperature with 0.01M phosphate-buffered saline (PBS. pH=7.4) before and after each incubation step. Sections were incubated for 48h at 4°C in a humid chamber in PBS containing 0.3% Triton X100 (Sigma), 5% normal goat serum (Sigma), and a cocktail of primary antibodies made of a polyclonal rabbit anti-KP (1:12000; #564; gift from A. Caraty) and a monoclonal mouse anti-BEND (1:1500; MAB5276, Chemicon, Billerica, MA, USA). For detection, sections were incubated for 90 min at room temperature with biotinylated goat anti-rabbit IgG (1:500; Vector, Burlingame, CA, USA) followed by a mixture of fluorescein-linked streptavidin (1:500; Vector) and Texas red linked goat anti-mouse IgG (1:300; Jackson Immunoresearch, West Grove, PA, USA) for 1h. Sections were then washed with PBS and coverslipped using Vectashield (Vector). The specificity of the polyclonal antibody raised against KP has been validated previously in sheep (15). Specificity of the monoclonal mouse anti-BEND has been checked by preadsorbing the antibody with an excess of synthetic BEND (10µg/mL - Genecust, Dudelange, Luxembourg) for 24h at 4°C, followed by immunocytochemical procedure as described above, which resulted in the absence of immunoreactivity.

*Analysis of results*

Ten sections taken 140 µm apart from the medial to the caudal part of the ARC have been examined for each animal using a confocal microscope (Zeiss LSM 700, Oberkochen, Germany). Counting was performed by an experimenter unaware of the photoperiod status of the animal. The microscope was equipped with a 40x objective and with 488 nm and 555 nm diode laser lines. Ten KP neurons and 10 BEND neurons were randomly selected in each animal and images were acquired. All images were then captured as 7-12 µm thick z-stacks with a 1 µm gap between images in sequential mode, using dichroic mirrors to avoid altering fluorescence intensity and any "bleed-through". Twenty-four-bit confocal images were acquired with a frame size of 1.024 x 1.024 pixels. Z-stacks were subsequently examined and the number of appositions per neuron was then estimated for each animal in both experimental groups. KP apposition on BEND neurons has been quantified when KP immunoreactivity was observed in contact with a BEND immunoreactive neuron. Inversely, BEND apposition on KP neurons has been quantified when BEND immunoreactivity was observed in contact with a KP immunoreactive neuron. The mean number (± S.E.M.) of appositions per neuron was then calculated for 10 KP and 10 BEND neurons per animal and statistical comparison between the two groups of ewes (exposed to LD vs SD) have been performed using the Mann-Whitney U test after checking that data of our two independent groups were not normally distributed (GraphPad Prism software, release 6.07).

**Exp 2 – Does BEND activates KP neurons in the ARC?**

*Light treatment and stereotaxic approach*

Ewes (n=9) were synchronized with LD for 75 days and transferred to a SD light regimen. After 45 SD, ewes were implanted with a cannula in the third ventricle under gaseous isoflurane anesthesia, with their heads placed in a stereotaxic frame. The skull was exposed and the stereotaxic frame was aligned at bregma using visual landmarks. Using ventriculograms for guidance, the cannula was placed into the third ventricle. The second set of X-rays was done to confirm the correct placement of the cannula before fixation on the skull. Ewes were allowed to recover for 6 weeks until their LH levels were >1ng/mL for three consecutive blood samples before injection of the peptide. Ewes were divided into 2 groups of animals either receiving the BEND solution (5 µg in 100 µL saline; Genecust; dose chosen from previously published work: (4); n=5) or 100 µL vehicle through the cannula (n=4). Two hours after injection, ewes were sacrificed and brains were collected after perfusion as described in the "Brain sampling" section.

***KP/c-Fos immunocytochemistry.*** Sections were incubated in PBS containing 0.3% Triton X100 (Sigma), 5% normal goat serum (Sigma), and the polyclonal rabbit anti-KP (1:12000, #564; gift from A. Caraty) in combination with a monoclonal mouse anti-c-Fos (1:5000, sc-271243, Santa Cruz, Dallas, TX, USA) for 48h at 4°C in a humid chamber. For detection, sections were incubated for 90 min at room temperature with biotinylated goat anti-rabbit IgG (1:500; Vector) followed by a mixture of fluorescein-linked streptavidin (1:500; Vector) and Texas red linked goat anti-mouse IgG (1:300 ; Jackson Immunoresearch) for 1h. Sections were then washed with PBS and coverslipped with Vectashield (Vector). The mouse monoclonal c-Fos antibody was previously validated for use in sheep brain (35).

*Analysis of results.*

Twelve representative microscopic fields obtained on 12 sections taken 140 µm apart ranging from the rostral to the caudal ARC have been analyzed for each ewe. Pictures have been taken for each section using an Olympus BX51 microscope equipped with a U-MNIBA2 filter (Olympus, Tokyo, Japan) allowing detection of FITC-labeled neurons (excitation wavelength: 470‑490 nm) and with a U-MWIG2 filter (excitation wavelength: 520‑550 nm) for visualization of Texas Red-stained cells. Doubled-labeled neurons were discernible by switching from one filter block to the other or by merging the images obtained under each illumination with a digital camera using SPOT Advanced software (Digital Instruments, Burroughs, MI, USA). KP neurons and KP/c-Fos double-labeled neurons were quantified manually by an investigator who was blind to the group allocation. Numbers (±SEM) of KP neurons/mm$^2$, c-Fos neurons/mm$^2$, and double-stained neurons/mm$^2$ were then calculated for each ewe as well as the percentage of activated KP neurons and the percentage of c-Fos neurons expressing

KP. Statistical comparison between the two experimental groups (BEND or control ewes) was performed using a Mann-Whitney U test after checking that data of our two independent groups were not normally distributed (GraphPad Prism software, release 6.07).

**Results**

*BEND neurons appositions on KP neurons are modulated by photoperiod*

The distribution of immunoreactive KP and BEND neurons is similar to that previously reported in this species (6, 15) at the level of the ARC and overlap mainly in the caudal two-thirds of the structure. However, no colocalization is observed. Consistent with previous reports (6, 20, 21), the density of KP neurons is greater in SD-exposed ewes than in LD-exposed ewes while BEND labelling appears more intense under LD. In both groups of ewes, numerous KP buttons can be observed near BEND somas (Fig. 1 A, top panel) as well as a dense network of BEND fiber terminals apposed to KP neurons (Fig. 1 A, bottom panel). Confocal z-stack images analysis revealed a similar number of KP synaptic buttons on BEND neurons between LD (17.82±3.26 buttons/neuron) and SD (22.20±1.31 buttons/neuron) ewes (Fig. 1B, top set of box plots). In contrast, the number of BEND appositions on KP neurons was significantly smaller in LD (25.96±3.59 buttons/neuron) than in SD (44.32±2.88 buttons/neuron) animals ($p=0.004$, Mann-Whitney U test; median values: LD 27.30, SD 42.40; Fig. 1 B, bottom set of box plots).

*Intracerebroventricular BEND injection activates KP neurons*

To determine if BEND affects KP neuron activity, SD ewes were implanted with cannulae and injected into the third ventricle with 5 µg of peptide for 2h before being killed. Activation of KP neurons was then assessed by seeking early gene activation marker c-Fos (Fig. 2A-C). The occurrence of an activated neuron was easily discernible as both immunoreactivities occupy distinct cell compartments (nucleus for c-Fos and cytoplasm for KP, Fig. 2C). Activated c-Fos neurons were observed both in BEND and control animals but they were statistically more numerous in BEND injected animals (controls: 402±25 neurons/mm$^2$ vs. BEND: 663±28 neurons/mm$^2$ ; Mann Whitney U test; median values: control 402,7 and BEND: 632,0; $p<0.05$, Table 1). The number of KP neurons was also significantly greater in BEND treated ewes (274±14 neurons/mm$^2$) than in control animals (210±4 neurons/mm$^2$) as well as the number of double-labelled neurons (controls :20±2 neurons/mm$^2$ vs. BEND: 45±5 neurons/mm$^2$ ; Mann Whitney U test; median: control 19.41 and BEND: 40.63; $p<0.05$). The percentage of c-Fos neurons immunoreactive for KP did not vary significantly between groups (5.02% in controls vs 6.73% in treated animals) while the percentage of activated KP neurons was significantly greater in BEND-injected ewes (16.26% vs. 9.35%; p<0.05, Mann Whitney U test; median values: BEND-treated group 15.10, Control group 9.06; Table 1).

**Discussion**

Our study revealed photoperiod-associated variations of the number of BEND terminals on KP neurons from the ARC in ewes. We observed a significant increase when the gonadotropic axis is activated. We also demonstrated a selective activation of ARC KP neurons 2 hours after a 5µg BEND infusion into the third ventricle in ewes with an active gonadotropic axis.

The results obtained in the present study showing significant plasticity of BEND contacts on KP neurons with photoperiodic status add to previous data showing seasonal remodeling of BEND appositions on GnRH neurons (7) and the close association of the synaptic plasticity marker PSA-NCAM with BEND somas of the ARC (8). Collectively, these findings support a role for BEND neurons in plasticity processes, which may be involved in the seasonal control of breeding. The study of Jansen et al (7) reported complex seasonal regulation of the number of BEND appositions both on GnRH somas and dendrites depending on the GnRH subpopulations (rostral and caudal distributions) in an ovariectomized estradiol-replaced sheep model. A larger number of BEND appositions was found on somas of both GnRH subpopulations during anestrus. However, the situation for GnRH dendrites was more complex as the number of BEND appositions on GnRH dendrites from the caudal subpopulation of GnRH neurons, likely poorly involved in episodic GnRH/LH secretion, increased specifically during the breeding season. This latter finding is reminiscent of the greater number of BEND appositions on ARC-located KP neurons in SD ewes we report here. Overall, the differences in the response may stem from different seasonal and local feedback regulation exerted through sex steroid receptors, estrogen receptor α (12) and progesterone receptor (13), expressed within BEND neurons in sheep but may also originate from the involvement of BEND in the transmission of metabolic or stress cues (36).

In contrast, the number of KP terminals on BEND neurons did not vary with the photoperiodic treatment, which suggests that seasonal remodeling of appositions at the surface of BEND neurons does not include those arising from KP neurons. The increased number of BEND appositions on KP neurons during SD may appear counterintuitive as KP is a potent stimulator of GnRH/LH secretion while BEND was reported to inhibit LH secretion. The inhibitory action was observed in follicular phase ewes (9) but also in other mammalian species belonging to different clades such as rats (37), pigs (38) and primates (39). We also observed that BEND infusion induces an activation of KP neurons, as shown by an increase in c-Fos expression. This raises the possibility that BEND may act directly or indirectly on KNDy neurons to affect the basal level of GnRH secretion. A similar hypothesis was proposed for somatostatin, an inhibitory modulator of LH release (40), whose secreting neurons in the ewe ARC are activated during the LH surge (41). Another hypothesis is that BEND neurons will provide part of the negative regulation of the "pulse generator" associated with GnRH/LH pulsatile release. This would take place through the activation of dynorphin synthesis produced in KNDy neurons as dynorphin has an autocrine inhibitory role on KP release. Indeed, it was shown in the ovine ARC that dynorphin activation precedes that of KP and neurokinin B, which would be consistent with our hypothesis (42). Interestingly, in the same article, Fergani and colleagues also demonstrated that activation of BEND neurons during follicular phase appears 16 hours after progesterone withdrawal while dynorphin activation was maximal just before activation of sexual behavior (which appeared 28.5 hours after progesterone withdrawal). Furthermore, activations of KP and neurokinin B were significantly elevated around surge time i.e. 36.7 hours after progesterone withdrawal (42). From these data, we may hypothesize that BEND treatment may have elicited an increased activation of dynorphin secretion within KP neurons as 94% of KP neurons in the ARC express dynorphin (26).

As BEND neurons of the ARC project towards GnRH neurons located rostrally at the level of the OVLT (13), it may be that the negative action on LH secretion occurs only in the vicinity of GnRH somas as suggested by the work of Goodman and collaborators (11), and that appositions on KP cell bodies in the ARC transduce only metabolic status in line with the hypothesis raised previously by Jansen and coworkers regarding the role of BEND terminals on GnRH dendrites (7). Of note, we allowed 2 hours between BEND injection in the third ventricle and animal killing since the maximum level of c-Fos concentration is reached after ~90 minutes following cellular stimulation (43), this time may have allowed the occurrence of interneurons projecting on KP neurons to be activated by BEND, therefore, we cannot exclude that expression of c-Fos by KP neurons resulted from both direct and indirect effects of BEND. Indeed, as BEND was infused in the third ventricle, the peptide could also have diffused into the ventricular system and reached cells expressing the µ opioid receptor located outside the ARC in areas close to the third ventricle such as the preoptic area or the dorsomedial nucleus (44). Among cellular populations expressing the µ opioid receptor, GABA-producing neurons may be good candidates since they are widely spread at the level of the preoptic area (45, 46). In addition, GABAergic inputs on ovine KNDy neurons have been observed (47). As BEND is also involved in stress-associated suppression of reproduction in several species (48, 49), BEND central infusion may have activated some stress-related neuronal networks known to project on KNDy neurons such as the CRH network. Indeed, infusion of the µ opioid receptor antagonist, β-funaltrexamine, blocks stress-associated LH suppression induced by electric footshocks in rats (48). Whether or not BEND may involve CRH neurons to modulate LH release in sheep remains to be established but in contrast with other species, it was previously shown in steroid-replaced ewes that central administration of 20 nmol of CRH stimulates LH release (50), therefore the stimulation of KP neurons following BEND infusion may pass through activation of CRH release that in turn will activate KNDy neurons but as stated in a recent review (51), CRH effect on LH release may be extremely complex in the ovine species as it depends on the gonadal status and the time of the ovarian cycle.

BEND binds mainly µ opioid receptors whose distribution in the ewe mediobasal hypothalamus does not vary with the stages of the estrus cycle in the ewe (44). However, a recent article reporting the use of single cell RNA-Seq on mouse hypothalamic cells reported that KP neuronal subtypes do not express δ opioid receptors, often express κ and scarcely express µ opioid receptors (46). This finding was extended in ovariectomized rats with low estradiol levels using a double in situ hybridization technique, which revealed that µ opioid receptor mRNA was virtually absent from KP neurons of the ARC (52). Therefore, any direct effect of BEND on KP neurons of the ARC in rodents seems unlikely. Whether this applies to the ewe remains unknown. As stated earlier, an indirect action of BEND upon KP neurons of the ARC through neurons expressing µ opioid receptors is indeed plausible. POMC neurons in addition to BEND also produce α-melanocortin stimulating hormone (MSH) known to reduce food intake and stimulate sexual behavior in rodents (53) by binding to MC3R and MC4R receptors. Of interest, using RNAscope technique these two receptors have been recently identified in KP neurons from the ram arcuate nucleus (54). Therefore, it may be that the increased number of BEND appositions on KP neurons we observed during SD may allow for an increased influence of αMSH on KP neurons that express MC3R and MC4R. However, a recent report pinpointed the lack of effect of αMSH on KP neuron excitability both in adult male and female mice (55) suggesting that αMSH is not involved in the control of GnRH/LH secretion, or at least not directly. Whether or not this applies to the sheep remains to be investigated.

Among the neuronal populations that may relay POMC neuron influence, some evidence gained in rats pinpoints towards dopaminergic neurons that may be sensitive to BEND effect since they express µ opioid receptor (14) and it was demonstrated in sheep that, depending on season, between 40% and 80% of KNDy neurons express D2R dopamine receptors (56, 57). Moreover, an electrophysiological study in transgenic mice performed by Zhang and van der Pol (58) demonstrated an inhibition of bursting electrical activity in ARC dopamine neurons when a µ opioid receptor antagonist is applied, which suggests direct activation of this neuronal population by BEND. As the effect of dopamine on ARC KP neurons seems to be mostly inhibitory (56), this population appears as a good candidate to transduce inhibitory BEND effects on neurons responsible for the timely release of LH. Recent work in rats also suggested that GnRH/LH inhibition under BEND influence may involve, at least in part, ARC glutamatergic neurons as double *in situ* hybridization revealed that approximately 10% of these neurons express µ opioid receptors (52) and that NR1, a subunit of the N-methyl-D-Aspartate (NMDA)-type glutamate receptor, is expressed in most ARC KP neurons in female rats (59).

Histochemical analysis also showed an increase of glutamatergic appositions on ARC KP neurons at the time of LH surge in the ewe reinforcing that hypothesis (60). This number of glutamatergic appositions on KNDy neurons depends on estradiol levels as it was at its highest with estrogen levels similar to those in the follicular phase (61).

This work demonstrated that BEND neurons are involved in seasonal plasticity at the level of the ARC where the number of BEND terminals on KP neurons is greater when the gonadotropic axis is active. Surprisingly, BEND infusion augments KP neuron activation, a process that may involve both direct and indirect effects. Since KP neurons may not express the µ opioid receptor, an indirect impact is privileged. This may involve distinct neuronal hypothalamic populations, which are known to express µ opioid receptors and to affect KP neurons in sheep and other species. Future work will be required to determine if these populations are also involved in BEND effect in sheep and should focus on elucidating the functional significance of this activation for the central control of reproduction.

**Acknowledgments**

The authors thank the staff of UEPAO and CIRE facilities for taking good care of our animals pre- and post-surgeries. We also thank the staff of PIC platform for maintenance of confocal microscope equipment and Ms. Stéphanie Martinet for her help in plasma collection. This research did not receive any specific grant from any funding agency in the public, commercial or not-for-profit sector.

**Data availability statement**

Images, data and information/reagents are available upon reasonable request.

**Table 1: Quantitative analysis of BEND effect on KP neuronal activation**

| Animal treatment | KP/mm² | c-Fos/mm² | Double-labeled neurons / mm² | % Double-labeled neurons/KP | % Double-labeled neurons/c-Fos |
|---|---|---|---|---|---|
| BEND (n=5) | 274 ± 14* | 663 ± 28* | 45 ± 5* | 16.26 ± 1.61* | 6.73 ± 0.70 |
| Control (n=4) | 210 ± 4 | 401 ± 25 | 20 ± 2 | 9.35 ± 1.27 | 5.02 ± 0.87 |

*Mann-Whitney U test: p<0,05, treated vs. control*

**Figure legends**

**Figure 1: (A)** Microphotographs of frontal sections at the level of arcuate nucleus of LD and SD treated ewes after immunolabelling directed against BEND (red staining from Texas red) and KP (green staining from FITC). Note the occurrence of KP terminals in close proximity of BEND (top panel) and of BEND terminals on KP (bottom panel) neurons respectively in both photoperiodic conditions. Scale bar= 5 µm. **(B)** Box plots showing the mean number (±SEM) of apposition per BEND (top panel) and KP (bottom panel) neurons under LD (white box plots) or SD (grey box plots) status. **p<0.005, Mann Whitney U test.

**Figure 2:** Microphotographs of representative frontal sections at the medial level of the arcuate nucleus after a double immunocytochemical labelling to detect KP (A-B; yellow green fluorescence from FITC) and c-Fos (C-D; red fluorescence from Texas red) following either vehicle (A,C,E) or BEND (B, D, F) injection into the third ventricle. Double-labelled neurons (arrowheads) can easily be seen on a merged image (E and F) for the different intracellular localization of the two epitopes. Scale bars= 50 µm in A-F and in inserts. Box plots on the right panel show the mean number of KP neurons/mm$^2$ (G), mean number of c-Fos neurons/mm$^2$ (H) and the mean percentage of double-labeled neurons in relation with KP neurons (I) in control (white box plots) and BEND-treated (grey box plots) groups respectively.